# Designation of Intra-layer and Intercalated High-Entropy Quasi-2D Compounds

Hong Xiang Chen[*, #, a, b, c], Sheng Li[#, a, b], Shu Xian Huang[a, b], Li An Ma[a, b, c], Sheng Liu[a, b], Fang Tang, Yang Chen, Yong Fang[*], Pin Qiang Dai[*, a, b]

[a] School of Materials Science and Engineering, Fujian University of Technology, Fuzhou 350118, China

[b] Fujian Provincial Key Laboratory of Advanced Materials Processing and Application, Fuzhou 350118, China

[c] Fujian Provincial Key Laboratory of Quantum Manipulation and New Energy Materials, Fuzhou, Fujian 350118, China

[d] Center for Advanced Energy and Functional Materials, Fujian University of Technology, Fuzhou 350118, China

[e] Jiangsu Laboratory of Advanced Functional Materials, Department of Physics, Changshu Institute of Technology, Changshu 215500, China

Email: hungxchen@163.com  (H. X. Chen), fangyong@cslg.edu.cn (Y. Fang), pqdai@126.com (P. Q. Dai);

**Abstract:** Here, we designed two promising schemes to realize the high-entropy structure in a series of quasi-two-dimensional compounds, transition metal dichalcogenides (TMDCs). In the intra-layer high-entropy plan, (HEM)$X_2$ compounds with high-entropy structure in the $MX_2$ slabs were obtained, here HEM means high-entropy metals, such as TiZrNbMoTa. And superconductivity with a $T_c$~7.4 K was found in a Mo-rich HEM$X_2$. On the other hand, in the intercalation plan, we intercalated HEM-atoms (FeCoCrNiMn) into the gap between the sandwiched-$MX_2$ slabs resulting a series of (HEM)$_x$M$X_2$ compounds, $x$ in the range of 0~0.5, in which HEM is mainly composed of $3d$ transition metal elements, such as FeCoCrNiMn. As the introduction of multi-component magnetic atoms, ferromagnetic spin glass states with strong 2D characteristics ensued. Tuning the $x$ content, three kinds of two in the high-entropy intercalated layer were observed including the $1 \times 1$ triangular lattice and two kinds of superlattices $\sqrt{3} \times \sqrt{3}$ and $\sqrt{3} \times 2$ in $x$=0.333 and x>0.5, respectively. Meanwhile, the spin frustration in the two-dimensional high-entropy magnetic plane will be enhanced with the development of $\sqrt{3} \times \sqrt{3}$ and will be reduced significantly when change into the $\sqrt{3} \times 2$ phase. The high-entropy TMDCs and versatile two-dimensional high-entropy structures found by us possess great potentials to find new physics in low-dimensional high-entropy structures and future applications.

**Introduction**

With the flourish of two-dimensional materials started with the discovery of graphene[1–4], quasi-two-dimensional compounds[5,6] which can be easily exfoliated to single or few-layered have been paid lots of attention, especially transition metal dichalcogenides (TMDCs). TMDCs are a big family of compounds with a chemical formula $MX_2$, in which M=transition metals (Ti, V, Nb, Ta, Mo, Hf, and so on), X=chalcogen elements (S, Se, Te). TMDCs have not only versatile chemical compositions but also various polymorphs[7], such as 1T, 2H, 3R, and so on. Various famous materials with exotic physical properties such as extreme magnetoresistance[8], superconductivity, type-II Weyl fermions[9–11], quantum spin Hall effect[12], and excellent application prospects such as electro- or photo- catalysis, photo-electronics[13][14], energy storage, pollution reduction[15], were found in TMDCs. These compounds show excellent structural compatibility which can be doped in both M and X sites[16]. On the other hand, transition metal metals, alkali metal metals, alkali earth metal atoms, rare-earth metal atoms, organic molecules can also be intercalated into the vacancies between two X-atoms layers from two $MX_2$ slabs to form intercalated compounds[17–19] such as $Cu_xTiSe_2$[20], $(NH_3)LiNbSe_2$ and so on.

Recently, high entropy structuralizing was found to be an effective way to search new materials. Since the idea of high-entropy alloy proposed by Ye[21,22], more than 400 kinds of high-entropy alloys were found until now. Since 2015, first high-entropy oxide (MgCoNiCuZn)O was found by M. Rost et al. [23], more attention was paid to high-entropy compounds. At present, the family of high-entropy compounds has been expanded to borides [24], oxides [23,25,26], carbides[27,28], sulfides[29], silicide[30] and so on. Some of them show excellent properties even surpassing those of the constituent materials[31], such as colossal dielectric constant [32], ultra-low thermal conductivity [29,33], enhanced mechanical [34], energy-storage ability [35] and so on. On the other hand, how the high-entropy effect on superconductivity [36–42] and magnetic ordering [25,26,28,43–45] are quite concerned in condensed matter physics. Although so many high-entropy compounds have been found, the high-entropy compounds in low-dimensional are still in rare.

Considering the versatile chemical compositions and structure compatibility, TMDCs could be a promising system to realize the high-entropy two-dimensional (2D) or quasi-2D materials. In this work, we found two feasible ways to realize high-entropy structures in TMDCs. One way is to realize high-entropy structure at the M site, and the other is to intercalate high-entropy metal-atoms into the intra-layer vacancies between MX2 slabs. For intralayer scheme, we found both equiatomic and non-equiatomic $(HEM)X_2$ compounds including a six component $(HEM)X_2$, $(TiVHfNbTaMo)Se_2$. In the intercalation scheme, we successfully intercalated FeCoCrNiMn/FeCoCrNi simultaneously into the gaps between $TiS_2$ or $TiSe_2$ slabs. The structural evolution with the intercalation content were systematically studied. The upper limit of intercalation content is around 0.5. Meanwhile, millimeter-sized single crystals of high-entropy TMDCs were grown by a chemical vapor transport method. Three kinds of crystal structures in $(HEM)_xTiX_2$ were revealed by X-ray and electron diffraction results. Electrical and magnetic properties of the high-entropy quasi-2D compounds were characterized. Surprisingly, superconductivity with a $T_c$ = 7.4 K was observed in a Mo-rich high-entropy compound, and ferromagnetic spin-glass states with strong anisotropy were found. And the x-content dependence of the magnetic transitions and spin frustration were systematically studied. The two high-entropy schemes we proposed can also be applied to other quasi-2D or quasi-1D systems.

**Results**

1T and 2H phases are the most commonly seen polymorphs in TMDCs. Here we take 1T phase as an example to illustrate the two schemes to realize high-entropy structures in TMDCs. In the single $MX_2$ slab, one layer of M atoms was sandwiched by two X-atom layers, and one M atom is octahedrally coordinated with six X-atoms. $MX_2$ slabs are mainly van-der-Waals connected with each other which can be easily exfoliated to single- or few- layered. Replacing M site with HEM elements, we will get (HEM)$X_2$ which owns a high-entropy structure inside the $MX_2$ slabs, as shown in Figure 1(c).

Searching for the possible combination of HEM elements, we found two equiatomic (HEM)$X_2$ in 1T phase, $(Ti_{0.2}V_{0.2}Zr_{0.2}Nb_{0.2}Ta_{0.2})Se_2$, $(Cr_{0.2}V_{0.2}Zr_{0.2}Nb_{0.2}Ta_{0.2})Se_2$ and $(Ti_{0.2}Zr_{0.2}Nb_{0.2}Mo_{0.2}Ta_{0.2})Te_2$, See the XRD results of (HEM)$X_2$ in Figure S1. The X-ray diffraction results indicate the pure 1T phase of the as-prepared powder. Meanwhile, slight amount of Mo (~4%) can be doped into (TiVHfNbTa)$Se_2$ as observed in a six-component single-crystal grown by the CVT method, (TiVHfNbTaMo)$Se_2$. Similarly, single crystal of $(Ti_{0.23}Zr_{0.19}Nb_{0.35}Mo_{0.03}Ta_{0.27})Se_2$ were found, in which the Mo-content is only ~3%. As shown in Figure 1(a-c), the EDX analysis of the cleaved surface of as-grown (HEM)$X_2$ single crystals were given, which indicate the high homogeneity of all elements. Meanwhile, other non-equiatomic (HEM)$X_2$ single crystals, such as Mo-rich (HEM)$X_2$ compounds crystalline in 2H phase were found including $(Mo_{0.74}Nb_{0.15}Ta_{0.09}V_{0.02})Se_2$ and $(Mo_{0.65}Nb_{0.22}Ta_{0.10}V_{0.03})Se_2$ with the lateral size up to 1 cm were obtained.

In the other high-entropy scheme, we take advantage of the vacancies between $MX_2$ slabs. In 1T phase, it has octahedral vacancies surrounded by six X-atoms from two X-atom layers. As previous works reported, 3d transition metals including V, Fe, Co, Cr, Ni, Mn can be intercalated into the inter-layer gaps, form M'$MX_2$, M'=3d transition metals. The introduction of 3d metals will bring rich magnetic properties, such as anti-ferromagnetism, ferromagnetism and spin-glass. Here we think about an extreme situation. When several kinds of 3d metals were intercalated into the inter-layer vacancies simultaneously, they may all contribute to the magnetism and have different magnetic moments. Consequently, it may form a 2D layer with both high structural and magnetic entropy. At present, even co-intercalation of two kinds of 3d elements in TMDCs has never been reported. Therefore, we designed an intercalated high-entropy structure (HEM)$MX_2$, in which the HEM composed by the 3d transition metals. Here we use HEA to react with $MX_2$ at high temperatures, and surprisingly five kinds of 3d metal elements (FeCoCrNiMn) were successfully intercalated into $TiS_2$ and $TiSe_2$ simultaneously. Meanwhile single crystal of (HEM)$_x$$TiS_2$ and (HEM)$_x$$TiSe_2$ were obtained by CVT method using iodine as transport agent.

Figure 1(d) present the element mapping of EDX of an as-grown single crystal with nominal composition $(Fe_{0.2}Co_{0.2}Cr_{0.2}Ni_{0.2}Mn_{0.2})_{0.5}TiS_2$. The EDX-determined chemical composition is $(Fe_{0.22}Co_{0.22}Cr_{0.22}Ni_{0.21}Mn_{0.12})_{0.46}TiS_2$. The atomic ratio of Fe, Co, Cr, and Ni are almost equal, while the as-determined content of Mn is ~40 % less than the nominal composition. Figure 1(e) is the elements mapping of a single crystal in as-prepared $(Fe_{0.2}Co_{0.2}Cr_{0.2}Ni_{0.2}Mn_{0.2})_{0.333}TiS_2$ powder collected on an EDX equipped on a TEM, the as-determined composition is $(Fe_{0.20}Co_{0.21}Cr_{0.21}Ni_{0.19}Mn_{0.19})_{0.333}TiS_2$ which is almost equiatomic.

Using X-ray diffractions, we studied the crystal-structure evolution with the intercalation content $x$ of (HEM)$_x$$TiS_2$ and (HEM)$_x$$TiSe_2$, HEM=FeCoCrNiMn. The as-prepared $TiS_2$ and $TiSe_2$ are both in 1 T phase (space group: $P\bar{3}m1$. Figure 3 (a-c) shows the XRD patterns of (HEM)$_x$$TiSe_2$. As shown in Figure 3(a) and (b), when $x$ is not higher than 0.4, all peaks can be well indexed by $P\bar{3}m1$. Through Rietveld refinements, the lattice parameters were obtained as shown in Figure 3(d). The lattice parameter $c$ will decrease with the increasing $x$ when $x<0.2$, and then increase with increasing $x$. When $x=0.2$, $a = b = 3.5648(1)$ Å, $c = 5.9905(2)$ Å, in which $c$ decreased about 0.02 Å as compared with the one in TiS$_2$. It can be seen that around $x=0.35$, both $b$ and $c$ decreased suddenly. The lattice parameter $a$ almost increased with the increasing intercalant content. For $x=0.4$, $a = b = 3.5890(1)$ Å and $c = 6.0085(1)$ Å, comparing with the ones in TiS$_2$, $a$ increased about 0.05 Å.

When the intercalation content was increased to $x=0.45$, the diffraction peaks can no longer be indexed by the 1T phase but by a monoclinic phase with a $\sqrt{3}a_0 \times b_0 \times 2c_0$ superlattice. Similar structure evolution phenomena was also reported in other *3d* transition metal intercalation systems such as Fe$_x$TiSe$_2$[46–48]. (HEM)$_{0.45}$TiSe$_2$ crystalline in a body-centered space group of $I2/m$ (No.12) with the lattice parameters $a = 6.2468(4)$ Å, $b = 3.5885(2)$ Å, $c = 12.0022(5)$ Å, and $\beta = 89.570(4)°$. When $x = 0.45 - 0.55$, no obvious impurity was observed. Further increasing the intercalant content, several impurity peaks which can be indexed by a cubic lattice were observed. And the intensity of the impurity peak around 33° increased with the increasing x content, as shown in Figure 3(d). The lattice parameters of as-prepared monoclinic (HEM)$_x$TiSe$_2$ in nominal composition including $b = \frac{a}{\sqrt{3}}$ and $\frac{c}{2}$ were also summarized in Figure 1(e). Both, $b$ and $c$ increased with x increased. Through EDX analysis, we found that the furthering increasing of lattice parameters, should not be due to the increase of total intercalation amount of HEM atoms, but the increase of the Cr and Mn content among five HEM elements.

In (HEM)$_x$TiS$_2$, the critical boundary between monoclinic phase and hexagonal phase is around $x=0.50$. The 1T phase structure were remained when $x \leq 0.45$. Besides the diffraction peaks of (HEM)$_{0.5}$TiS$_2$ which can be indexed by $P\bar{3}m1$, when we carefully check the XRD pattern between $16° - 30°$, weak signal of superlattice can be observed as shown in the inset of Figure 3(g). Similar to the case in (HEM)$_x$TiSe$_2$ when *x* was increased higher than 0.6, the impurity appeared. Fig. 3(h) and (i) are the Rietveld refinement results of (HEM)$_{0.45}$TiS$_2$, the $R_p = 5.77\%$, $R_{wp} = 7.77\%$, and $R_{exp} = 4.46\%$ when $x=0.45$, and $R_p$=6.77%, $R_{wp}$=8.62%, and $R_{exp}$=4.73% suggest the reliability of as-solved structure. The refined lattice parameters are $a = b = 3.4271(1)$ Å and $c = 5.7832(2)$ Å for (HEM)$_{0.45}$TiS$_2$ ($P\bar{3}m1$) and $a = 5.9510(2)$ Å, $b = 3.4343(1)$ Å, $c = 11.5427(3)$ Å, and $\beta = 89.929(2)°$ for (HEM)$_{0.55}$TiS$_2$ ($I2/m$).

SAED patterns was collected to further study the superstructures in (HEM)$_x$TiS$_2$. Figure. 4 (a-c) are the SAED patterns of (HEM)$_{0.55}$TiS$_2$ which can be well indexed by the diffraction patterns of monoclinic phase ($C2/m$) collected from $<001>, <010>$, and $<11\bar{1}>$ zone axis, respectively, which are consistent with the XRD results. Surprisingly, we found another superlattice of $\sqrt{3}a_0 \times \sqrt{3}a_0 \times 2c_0$ in a (HEM)$_{0.333}$TiS$_2$ sample which was furnace-cooled from the sintering temperature (800℃). It should be noted that the results in Figure 3 are all from quenched samples. Comparing the XRD pattern of quenched and furnace-cooled samples (See Figure S2), we found the cooling rate shows significantly effect on the ordering in the HEM-atoms layer. The XRD pattern of furnace-cooled (HEM)$_{0.333}$TiS$_2$ can be well indexed by a $\sqrt{3}a_0 \times \sqrt{3}a_0 \times 2c_0$ superstructure and the space group was $P\bar{3}1c$ (No. 163). The refined lattice parameters are $a = b = 5.9071(7)$ Å and $c = 11.5073(1)$ Å, and refinement results are $R_p = 7.04\%$, $R_{wp} = 9.37\%$, and $R_{exp} = $

5.42 %. And the SAED pattern in Figure 4(e) can be well index by the <0001> zone. The atomic arrangement was observed in the HADDF image (Figure 4(f)), which matched well with the simulated crystal structure observed along the *c* axis.

Figure 4 (g-i) are the sketches of the atomic arrangements in single HEM-atom layer, the colorful balls are the HEM atoms. For sample with low intercalation content, the HEM-atoms will randomly occupied on the hexagonal lattice of the vacancies between $MX_2$ slabs as shown in Figure 4(g). When the intercalant content is around 0.5, the HEM atoms will be rectangularly arranged in the HEM plane. For sample with x=0.333, HEM atoms orderly occupied the sites of $\sqrt{3}a_0 \times \sqrt{3}a_0$ superlattice in which the triangular lattice was remained.

Transport properties of several $(HEM)X_2$ single crystals were characterized. And surprisingly we found superconductivity in a Mo-rich $(HEM)Se_2$ compound in a chemical composition $(Mo_{0.74}Nb_{0.15}Ta_{0.09}V_{0.02})Se_2$. The residual resistance ratio RRR=$R_{300K}/R_{7.5K}$=7.2, which indicates the high crystallinity of as-grown single crystal. As shown in Figure. 5(a) and (b), under zero field, the onset critical transition temperature $T_c^{onset}$ is about 7.4 K which is even slightly higher than the value 7.2 K in $NbSe_2$. And the superconducting transition finished at 7.06 K. As shown in Figure 4(c), the upper critical field $H_{c2}$ is about 2750 Oe. It should be noted that $MoSe_2$ is a semiconductor with an indirect bandgap of 1.08 eV[49]. To our best knowledge, no superconductor has ever been reported in Mo-site doped $MoSe_2$ or $MoS_2$. In contrast, no superconductivity was observed in the other Mo-rich $(HEM)Se_2$ single crystal $(Mo_{0.65}Nb_{0.22}Ta_{0.10}V_{0.03})Se_2$. It shows metallic behavior when *T*>17 K. However, below 17 K, an Anderson-localized *R-T* behavior was observed, which might be caused by the high-entropy structure in the HEM plane.

The magnetic properties of $(HEM)_xTiS_2$, *HEM*=FeCoCrNiMn, were shown in Fig. 6(a-d). As shown in Figure 6 (a), obvious irreversibility was observed between zero-field cooling (ZFC) and field cooling (FC) curves under 100 Oe. Here we define the peaks of ZFC curves as the spin-glass transition temperatures $T_g$. With the intercalant content *x* increasing, $T_g$ increased from 7.5 K (*x*=0.1) to 28.7 K (*x*=0.55). Similar behaviors were also observed in $(HEM)_xTiSe_2$, see Figure. S3. The isothermal magnetization curves up to 7 T measured at 2 K, as shown in Fig. 6(b), reveal the existence of ferromagnetism in $(HEM)_xTiS_2$. In the furnace-cooled sample, the coercivity is about 1150 Oe. And with $\sqrt{3}a_0 \times \sqrt{3}a_0 \times 2c_0$ superstructure developed when *x*=0.55, the hysteresis became more significant as shown in Figure. 6(b) in which the coercivity is up to ~6100 Oe. It should be noted that the magnetization behavior depends strongly to the loading method of sample due to the strong magnetic anisotropy. The isothermal magnetization curves measured at 40 K show paramagnetic behavior as shown in Figure S4.

$\frac{1}{\chi} - T$ curves and their Currie-Weiss fits of the data in the range of 200 K - 300 K, were shown in Figure 6(c). Using the Currie-Weiss fits, the Currie constants C, Currie-Weiss Constants $\Theta_{CW}$ were obtained as listed in Table. 1. The evolution of effective moment $\mu_{eff} = \sqrt{8C}$ with the *x* content and $T_g$ are summarized in Figure 6(d). Most samples own an effective moment in range of $2 - 3 \mu_B$. While for the samples of x=0.333, the as-fitted effective moment is much smaller than others, in which $\mu_{eff}$ equal to $1.22\ \mu_B$ and $1.08\ \mu_B$ in quenched and furnace-cooled sample, respectively. The spin frustration ratio, $f = |\Theta_{CW}/T_g|$, was found to be 0.99-4.29 for sample with $x \leq 0.3$. This indicates the moderate frustration occurred consistent with the crystal structure in which the HEM magnetic atoms are disorderly arranged in the triangular 2D-lattice with large number of vacancies. When x=0.333, the Currie-Weiss temperatures were increased significantly to -241.5 K and -385.3 K for quenched and furnace-cooled samples, respectively. Consequently, the frustration parameter was up to 29.5 in furnace-cooled sample. This reveals the strongly enhanced magnetic frustration with the $\sqrt{3} \times \sqrt{3}$ superlattice developed in the HEM plane. Different from the partially occupancy in the triangular 2D-lattice, the HEM atoms are fully occupied on the $\sqrt{3} \times \sqrt{3}$ superlattice. While for sample with $\sqrt{3} \times 2$ superlattice with x=0.55, the high entropy 2D magnetic lattice was change into a rectangular one, the frustration was significantly reduced as revealed by a relatively small frustration parameter $f = 0.26$. Therefore, ranking the spin frustration of three kinds of high-entropy magnetic lattice as shown in Figure 3(h, i, g), $\sqrt{3} \times \sqrt{3} > 1 \times 1 > \sqrt{3} \times 2$.

We also characterized the magnetic anisotropy in (HEM)$_x$TiX$_2$ using a single crystalline (HEM)$_{0.5}$TiSe$_2$ (in mass of 0.77 mg), with $H \parallel (001)$ and $H \perp (001)$, in which the EDX-determined chemical composition is (FeCoCrNiMn)$_{0.51}$TiSe$_2$. As shown in Figure 6 (e) and (g), the magnetization and the irreversibility between ZFC and FC are stronger for $H \perp (001)$ than $H \parallel (001)$. Comparing the isothermal magnetization curves at 2 K, the coercivity $H_c = 9950$ Oe of $H \perp (001)$ is much higher than the value 150 Oe in the case of $H \parallel (001)$. Similar strong in-plane and out-plane anisotropy was also observed in another (HEM)$_{0.2}$TiSe$_2$ single crystal as shown in Figure S6. The strong anisotropy reveals the 2D magnetic characteristics in (HEM)$_x$TiX$_2$.

Table 1: The Currie-Weiss fitted results of (HEM)$_x$TiS$_2$. C is the Currie constant, $\Theta_{CW}$ is the Currie-Weiss temperature, and $f = |\Theta_{CW}/T_g|$ is the spin frustration ratio.

| $x$ | C | $\Theta_{CW}$ (K) | $f$ |
|---|---|---|---|
| 0.1 | 1.059(8) | 7.4 | 0.99 |
| 0.2 | 0.672(2) | -53.0 | 4.29 |
| 0.3 | 0.733(2) | -47.1 | 3.23 |
| 0.333 | 0.185(1) | -241.5 | 16.48 |
| 0.333-super | 0.147(1) | -385.3 | 29.50 |
| 0.4 | 0.565(1) | -41.6 | 2.47 |
| 0.45 | 0.596(1) | -25.6 | 1.25 |
| 0.5 | 0.591(1) | -23.8 | 1.04 |
| 0.55 | 0.718(1) | -7.4 | 0.26 |

## Conclusion

In conclusion, we proposed two effective ways to realize the 2D high-entropy structures in TMDCs, which can also be applied in other quasi-2D or quasi-1D compounds. Various (HEM)$X_2$ with intra-layer high-entropy structure and (HEM)$_x$M$X_2$ with a high-entropy layer between M$X_2$ slabs were obtained here. Both samples with equiatomic and non-equiatomic compositions of HEM were found. Various large-size high-entropy single crystals were grown by the CVT method. Superconductivity with a $T_c$ up to 7.4 K was observed in a Mo-rich (HEM)$X_2$, (Mo$_{0.74}$Nb$_{0.15}$Ta$_{0.09}$V$_{0.02}$)Se$_2$. In (HEM)$_x$Ti$X_2$, we found the critical intercalant content is around 0.45-0.5 between $P\bar{3}m1$ and $I2/m$, in which HEM=FeCoCrNiMn/FeCoCrNi, and X=S/Se. What's more, when x=0.333, a $\sqrt{3}\times\sqrt{3}$ superstructure will be developed in (HEM)$_x$TiS$_2$. Three kinds of 2D high entropy magnetic-atom lattice were found. Spin-glass with ferromagnetic clusters were detected in (HEM)$_x$Ti$X_2$. The spin-glass transition temperature is up to 28.6 K. Strong spin frustration and reduced frustration were found in the $\sqrt{3}\times\sqrt{3}$ and $\sqrt{3}\times 2$ superlattice, respectively. And the strong magnetic anisotropy was detected in single crystals which reveal the 2D magnetic characteristics in (HEM)$_x$M$X_2$. Our work provides new ways to discover new high entropy compounds especially the intercalation method was first reported. And it will evoke more enthusiasm about exploring high-entropy compounds and the high-entropy effects in low-dimensional systems.

## Figures and Figure Captions

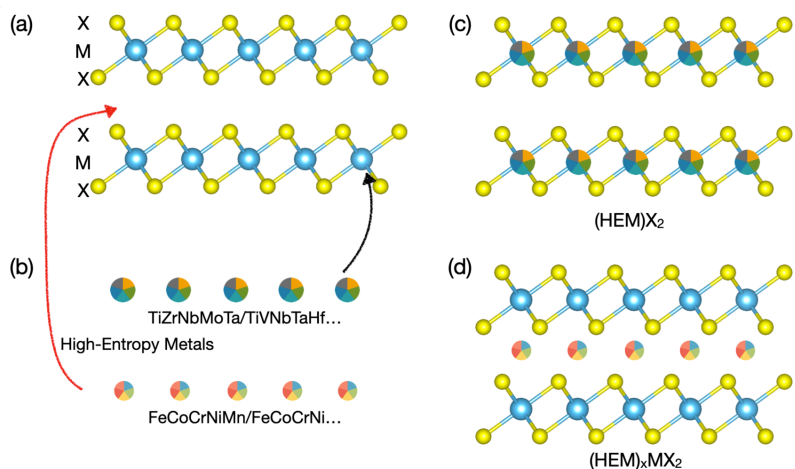

**Figure 1. Two schemes to realize high-entropy structures in MX$_2$.** (a) Crystal structure of M$X_2$ viewed from the $a$ axis, here we take 1T phase as an example. (b) High-entropy metal elements. (c) Crystal structure of (HEM)$X_2$. (d) Crystal structure of (HEM)$_x$M$X_2$.

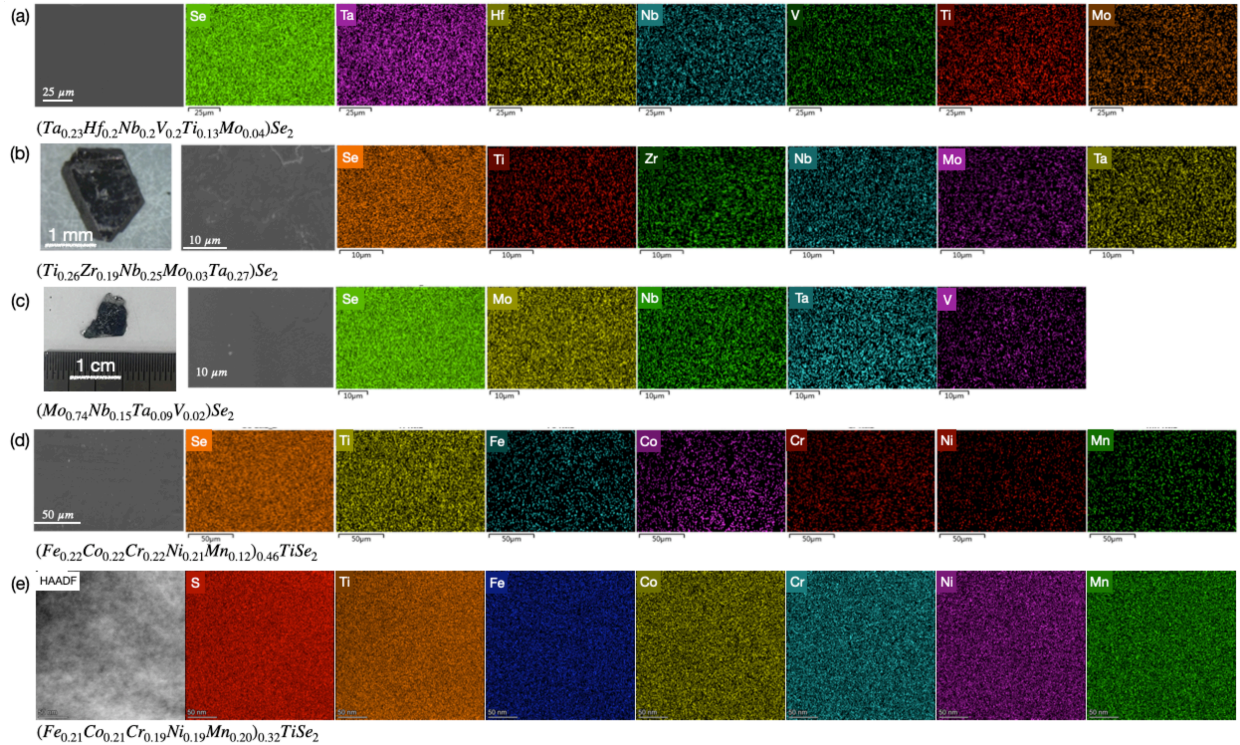

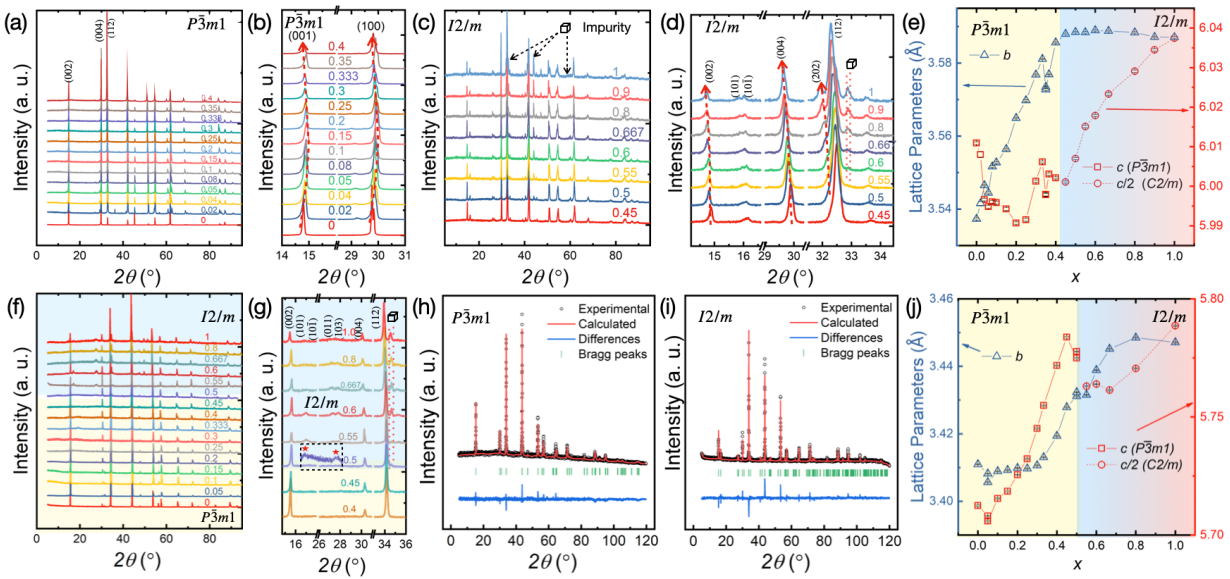

**Figure 2. Energy dispersion X-ray spectroscopy analysis of several (HEM)X$_2$ and (HEM)MX$_2$ compounds.** Element mapping analysis of (a) (TaHfNbVTiMo)Se$_2$, (b) (TiZrNbMoTa)Se$_2$, (c) (MoNbTaV)Se$_2$ and (d) (FeCoCrNiMn)$_{0.5}$TiSe$_2$ single crystals grown by CVT method. (e) Element mapping analysis of a (Fe$_{0.2}$Co$_{0.2}$Cr$_{0.2}$Ni$_{0.2}$Mn$_{0.2}$)$_{0.333}$TiS$_2$ collected by EDX on a TEM with high resolution. Other results are collected by the EDX on a FESEM.

**Figure 4. The X-ray diffraction analysis of (HEM)$_x$MX$_2$.** (a) and (b) are the XRD patterns of (HEM)$_x$TiSe$_2$ ($x$=0-0.4), in the range of 5°~90° and 14°~31°, respectively. The dashed lines indicate the change of peak-positions of (001) and (100). (c) and (d) are the XRD patterns of (HEM)$_x$TiSe$_2$ with $x$=0.45-1 in the range of 5°~90° and 14°~35°, the peaks of impurity are marked by arrows and cubic symbols. (e) The lattice parameters of (HEM)$_x$TiSe$_2$. (f) and (g) are the XRD patterns of Mn$_x$TiS$_2$ in the range of 5°~90° and 15°~36°, respectively. (h) Refinement results of (HEM)$_{0.45}$TiS$_2$. (i) Refinement results of furnace-cooled (HEM)$_{0.333}$TiS$_2$. (j) The refined lattice

parameters of (HEM)$_x$TiS$_2$.

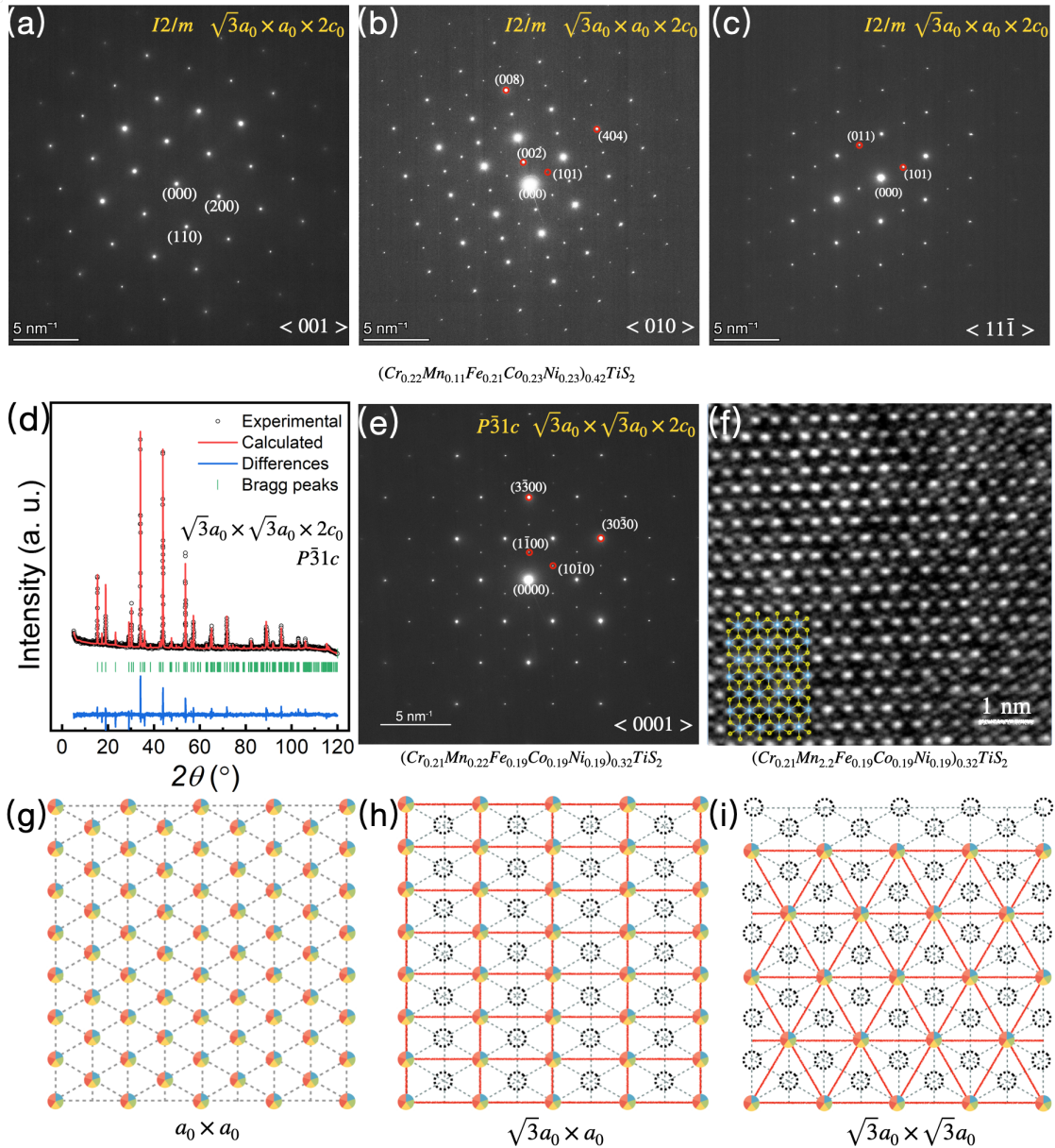

**Figure 4 The superstructures in (HEM)$_x$TiS$_2$.** (a), (b), and (c) are the SEAD **images of (**HEM)$_{0.5}$TiS$_2$ collected from $<001>$, $<010>$, and $<11\bar{1}>$ zone axis. (d) XRD refinement results of (HEM)$_{0.333}$TiS2 using a space group of $P\bar{3}1c$. (e) SAED image of (HEM)$_{0.333}$TiS$_2$, (f) HADDF image of (HEM)$_{0.333}$TiS$_2$. (g), (h), and (i) are the atomic arrangements in the 2D HEM plane in the structure of $a_0 \times a_0$, $\sqrt{3}a_0 \times a_0$, and $\sqrt{3}a_0 \times \sqrt{3}a_0$, respectively. The colored balls indicate the high entropy metal atoms, FeCoCrNiMn.

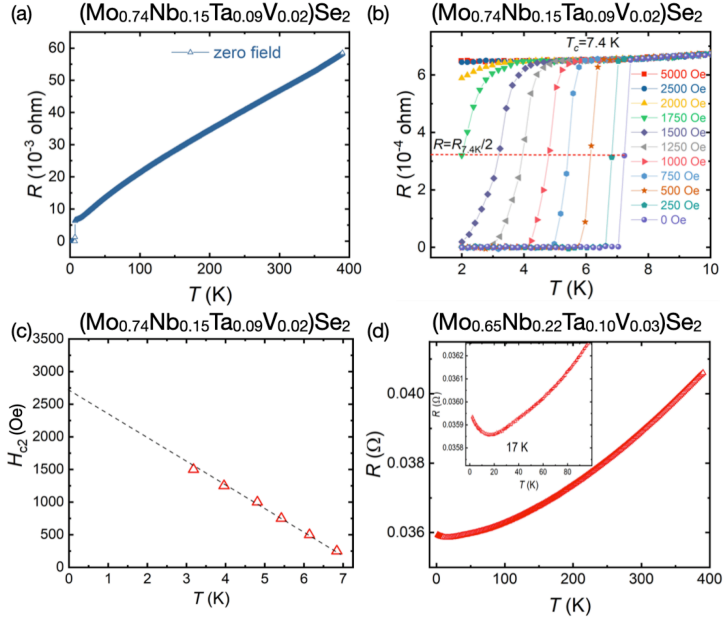

**Figure 5.** Transport properties of (MoNbTaV)Se$_2$. (a) Temperature dependent resistance of (Mo$_{0.74}$Nb$_{0.15}$Ta$_{0.09}$V$_{0.02}$)Se$_2$. (b) Field-dependent $R$-$T$ in the range of 1-10 K of (Mo$_{0.74}$Nb$_{0.15}$Ta$_{0.09}$V$_{0.02}$)Se$_2$. (c) Field-dependence of the upper critical field $H_{c2}$. (d) $R$-$T$ of Mo$_{0.65}$Nb$_{0.22}$Ta$_{0.10}$V$_{0.03}$Se$_2$.

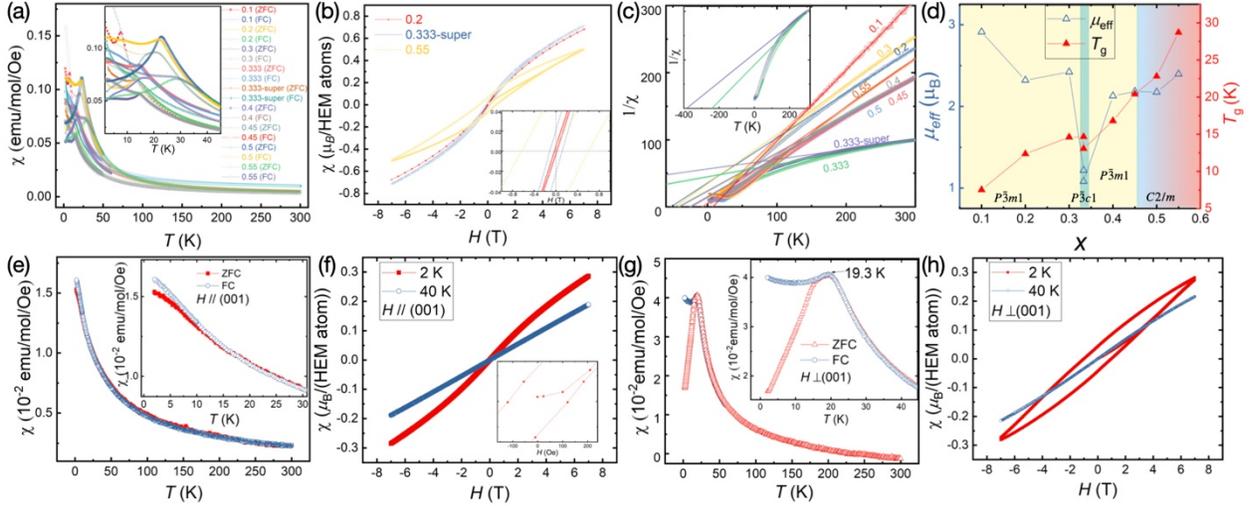

**Figure 6.** The magnetic properties of (HEM)$_x$TiX$_2$. (a) ZFC and FC curves of (HEM)$_x$TiS$_2$ measured under 100 Oe, (b) Isothermal magnetization of (HEM)$_x$TiS$_2$, (c) $\frac{1}{\chi}-T$ curves of (HEM)$_x$TiS$_2$, (d) x content dependence of spin-glass transition temperature ($T_g$) and effective moment of (HEM)$_x$TiS$_2$. (e) ZFC and FC curves of (HEM)$_{0.5}$TiS$_2$ single crystal measured under 2000 Oe with $H \parallel (001)$. (f) Isothermal magnetization of (HEM)$_{0.5}$TiS$_2$ measured under 2 K and 40 K with $H \parallel (001)$. (g) ZFC and FC curves of (HEM)$_{0.5}$TiS$_2$ single crystal measured under 2000 Oe with $H \perp (001)$. (f) Isothermal magnetization of (HEM)$_{0.5}$TiS$_2$ measured under 2 K and 40 K with $H \perp (001)$.

## Conflict of interest

The authors declare no interest conflicts in this work.

**Acknowledgments**
H. X. Chen would like to thank Dr. T. Ying, Dr. X. Sun, and Prof. X. Chen for help discussions, Dr. L. Zhao for magnetism measurements, Dr. Zhang from ZKKF (Beijing) Science & Technology Co., Ltd for TEM observations. This work was supported by the National Science Foundation for Young Scientists of China No. 51902055, Key University Science Research Project of Jiangsu Province of China (Grant No. 19KJA530003), and Open Fund of Fujian Province Key Laboratory of Quantum Manipulation and New Energy Materials (Grant No. QMNEM1902 and QMNEM1903,). Following institutes contribute equally in this work, School of Materials Science and Engineering, Fujian University of Technology, Fujian Provincial Key Laboratory of Advanced Materials Processing and Application, Fujian Provincial Key Laboratory of Quantum Manipulation and New Energy Materials.